\documentclass[%
 reprint,
superscriptaddress,
 amsmath,amssymb,
 aps,
 twocolumn,
float,prl]{revtex4-2}

\usepackage{graphicx}
\usepackage{dcolumn}
\usepackage{bm}
\usepackage{xcolor}
\usepackage[caption=false]{subfig}
\usepackage{tabularx}
\usepackage{epstopdf}

\begin{document}

\newcommand{\Schro}{Schr\"{o}dinger~}
\newcommand{\ket}[2]{\left|#1,#2\right\rangle} 
\newcommand{\bra}[2]{\left\langle#1,#2\right|} 
\newcommand{\me}[4]{\left\langle#1,#2\right|\tilde{\rho}\left|#3,#4\right\rangle} 

\title{Measurement of the Electron Magnetic Moment
}

\author{X. Fan}
 \email{xing.fan@northwestern.edu}
\affiliation{Department of Physics, Harvard University, Cambridge, Massachusetts 02138, USA}
 \affiliation{Center for Fundamental Physics, Department of Physics and Astronomy, Northwestern University, Evanston, Illinois 60208, USA}
 
 \author{T. G. Myers}
  \affiliation{Center for Fundamental Physics, Department of Physics and Astronomy, Northwestern University, Evanston, Illinois 60208, USA}
  
 \author{B. A. D. Sukra}
  \affiliation{Center for Fundamental Physics, Department of Physics and Astronomy, Northwestern University, Evanston, Illinois 60208, USA}
 
\author{G. Gabrielse}
 \email{gerald.gabrielse@northwestern.edu}
 \affiliation{Center for Fundamental Physics, Department of Physics and Astronomy, Northwestern University, Evanston, Illinois 60208, USA}
\date{\today}

\newcounter{gg}
\newenvironment{CompactEnumerate}{
\begin{list}{\arabic{gg}.}{\usecounter{gg} \setlength{\topsep}{-2pt} \setlength{\rightmargin}{30pt} \setlength{\itemsep}{-5pt}}}{\end{list}\smallskip}

\newcommand{\nuap}{\nu_a^\prime}
\newcommand{\nucp}{\nu_c^\prime}
\newcommand{\nuab}{\bar{\nu}_a}
\newcommand{\nucb}{\bar{\nu}_c}
\newcommand{\nuzb}{\bar{\nu}_z}
\newcommand{\numb}{\bar{\nu}_m}
\newcommand{\fcb}{\bar{f}_c}
\newcommand{\zhat}{\bf \hat{z}}

\begin{abstract}     
The electron magnetic moment,
$-\mu/\mu_B = g/2 = 1.001 \, 159 \, 652 \, 180 \, 59 \, (13) \, [0.13\, \rm{ppt}]$, 
is determined 2.2 times more accurately than the  value that stood for 14 years.  The most precisely determined property of an elementary particle tests the most precise prediction of the Standard Model (SM) to $1$ part in $10^{12}$. The test would improve an order of magnitude if the uncertainty from discrepant measurements of the fine structure constant $\alpha$ is eliminated since the SM prediction is a function of $\alpha$. The new measurement and SM theory  together predict $\alpha^{-1}=137.035 \, 999 \,
166 \, (15) \, [0.11 \, \rm{ppb}]$ with an uncertainty ten times smaller than the current disagreement between measured $\alpha$ values.
\end{abstract}

\pacs{13.40.Em, 14.60.Cd, 12.20-m}

\maketitle

The quest to find physics beyond the Standard Model of Particle Physics (BSM) is well motivated because the SM is  incomplete. No known $CP$ violation mechanism \cite{Sakharov}  is large enough to keep the matter and antimatter produced in the Big Bang \cite{ExpandingUniverse1946} from annihilating as the universe cooled, dark matter  \cite{DMDiscovery1933,DMRotationCurve1980} has not been identified, and neither dark energy \cite{Perlmutter_1999_AcceleratingUniverse,Riess_1998_AcceleratingUniverse} nor  inflation \cite{Linde1982Inflation,STAROBINSKY1980Inflation,STAROBINSKY1982Inflation,Guth1981Inflation} has a SM explanation. The most precise SM prediction is the electron magnetic moment in Bohr magnetons, $-\mu/\mu_B = g/2$, with $\mu_B=e\hbar/(2m)$ for electron charge $-e$  and mass $m$, and the reduced Planck constant $\hbar$.  It affords great BSM sensitivity \cite{ScreenedFifthForceLimitFromGFactor2018,HiggsDoubletCPViolation2019,Giudice2012,ProbeNewPhysicsWithElectronG2014,HiddenPhotonConstraintsHamaguchi,PhysRevD.98.075011,ALPExplanationOfAnomaly2020,ALightComplexScalarForTheElectronAndMuonAnomalousMagneticMoments,SUSYSenarioForElectronAndMuong_2019} in that BSM particles and electron substructure could shift the measured value from what is now predicted (analogous to how quark substructure shifts the proton moment). The SM sectors involved in the prediction include the Dirac prediction \cite{DiracTheoryOriginal},
quantum electrodynamics (QED) \cite{Schwinger,Petermann,Sommerfield:1958,Kinoshita:1995,LaportaRemiddi:1996,QED_C8_Lapo,QED_C10_nio,TenthOrderQED2012} with muon and tauon contributions 
\cite{KURZ2014_HeavyLeptons}, and also 
hadronic  \cite{Nomura2013,HadronicContribution2014,WeakHadronic}
and weak interaction contributions 
\cite{WeakCondtribution1,WeakCondtribution2,WeakCondtribution3,WeakCondtribution4}.  The SM prediction is a function of the measured fine structure constant, $\alpha$, displayed later in   Eq.~\ref{eq:SMTheory}.

\begin{figure}[htbp!]
\centering
    \includegraphics[width=\the\columnwidth]{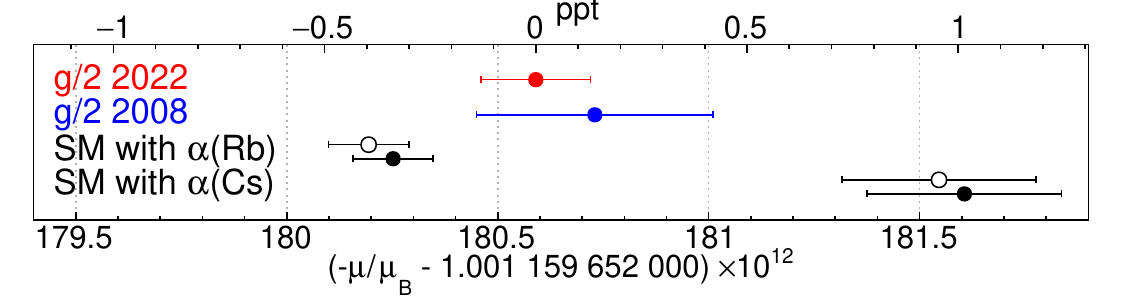}
   \caption{This Northwestern determination (red) and our 2008 Harvard determination (blue) \cite{HarvardMagneticMoment2008}.  SM predictions (solid and open black points for slightly differing  $C_{10}$ \cite{atomsTheoryReview2019,Volkov2019}) are functions of discrepant $\alpha$ measurements \cite{MullerAlpha2018,RbAlpha2020}. A ppt is $10^{-12}$.
   }  
   \label{fig:Result}
\end{figure}

A new measurement, carried out blind of any measurement or prediction, determines $\mu/\mu_B$ to $1.3$ parts in $10^{13}$ (Fig.~\ref{fig:Result}). Measured in a new apparatus in a lab at a different university, the new value is 2.2 times more precise than, and consistent with, the one that  stood for 14 years \cite{HarvardMagneticMoment2008}.  In the most precise confrontation of theory and measurement, the SM prediction agrees to 1 part in $10^{12}$.  Our determination and the SM calculation are precise enough for a test that is 10 times more precise, once the discrepancy in measured $\alpha$ values \cite{MullerAlpha2018,RbAlpha2020} is resolved.

A one-electron quantum cyclotron is used.  This is  essentially a single electron suspended in a magnetic field $\mathbf{B}=B\hat{z}$ and cooled to its lowest quantum states \cite{QuantumCyclotron}. The magnetic moment operator for a spin-1/2 electron,   
\begin{equation}
\boldsymbol{\mu}=- \frac{g}{2} \, \mu_B \, \frac{\mathbf{S}}{\hbar/2}, 
\end{equation}
is proportional to its spin $\mathbf{S}$, normalized to its spin eigenvalue $\hbar/2$.
The energy levels are 
\begin{equation}
E = h \nu_s m_s + h \nu_c (n + \tfrac{1}{2}),
\end{equation}
where $h=2\pi\hbar$. The cyclotron frequency is $\nu_c=eB/(2\pi m)$ and $n=0,1,...$. The spin frequency is $\nu_s = (g/2)\nu_c$ and $m_s=\pm 1/2$. In terms of these frequencies, and the anomaly frequency $\nu_a \equiv \nu_s-\nu_c$, 
\begin{equation}
-\frac{\mu}{\mu_B} = \frac{g}{2} = \frac{\nu_s}{\nu_c} =
1 + \frac{\nu_a}{\nu_c}. 
\label{eq:FreeSpacegOver2}
\end{equation}
An important feature of an electron measurement (not available with muons \cite{Muong-2PRL_2021}, for example) is that its cyclotron frequency can be measured {\it in situ}.  The electron thus serves as its own magnetometer insofar as $B$ cancels out of the frequency ratios. Choosing to measure $\nu_a/\nu_c$ rather than  $\nu_s/\nu_c$
(called making a $g-2$ measurement)
significantly reduces the effect of  frequency measurement uncertainties.  However, it does not evade or reduce at all the largest $g/2$ measurement correction and its uncertainty, as we shall see.   

\begin{figure}[htbp!]
\centering\includegraphics[width=\the\columnwidth]{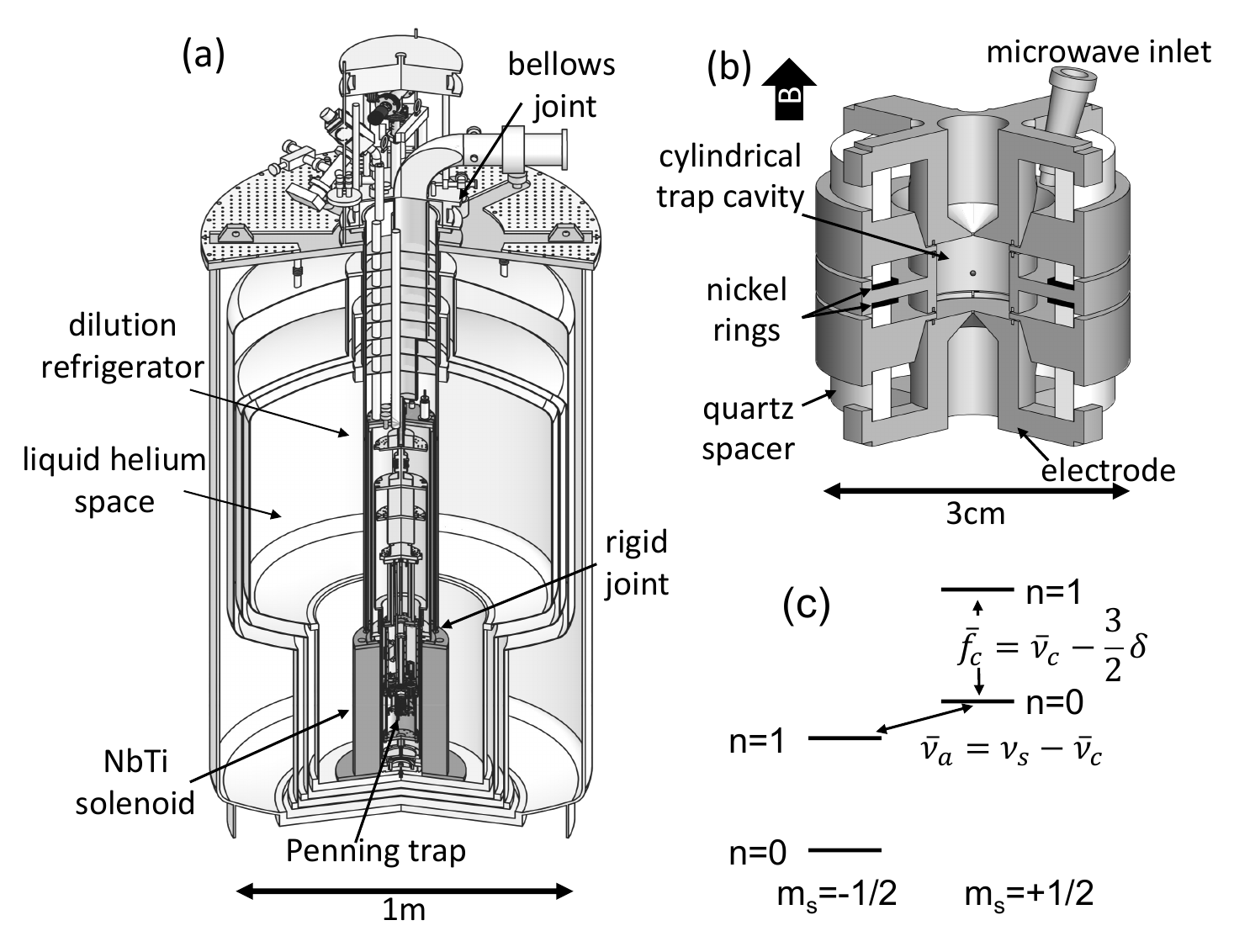}
  \caption{(a) Cryogenic system supports a 50 mK electron trap upon a 4.2 K solenoid to provide a very stable $B$.
  (b) Silver electrodes of a cylindrical Penning trap. (c) Quantum spin and cyclotron energy levels used for measurement.}
\label{fig:ApparatusAndLevels}
\end{figure}

A stable magnetic field is still critical at our precision because $\nu_a$ and $\nu_c$ are not measured simultaneously. Field drift of
2 ppb/day  \cite{Helium3NMR2019} (4 times below that in \cite{HarvardMagneticMoment2008})   makes  possible round-the-clock measurements, improved statistical precision, and better investigations of uncertainties.
The apparatus in Fig.~\ref{fig:ApparatusAndLevels}a achieves this 
by supporting a 50 mK trap on a 4.2 K self-shielding solenoid \cite{SelfShieldingSolenoid} using a mixing chamber flexibly hanging from the rest of a dilution refrigerator \cite{Atoms2019TowardImprovedMeasurement}.  

An electron in the field $B\hat{z}$ is trapped  by adding an electrostatic quadrupole potential $V \propto z^2 - \rho^2/2$, with $\boldsymbol{\rho} =
x\mathbf{\hat{x}}+y\mathbf{\hat{y}}$   \cite{Review}.
Cylindrical Penning trap electrodes \cite{CylindricalPenningTrap,CylindricalPenningTrapDemonstrated} (Fig.~\ref{fig:ApparatusAndLevels}b) with appropriately chosen relative dimensions and  potentials produce such a potential for a centered electron, which then oscillates nearly harmonically along $\hat{z}$ at the axial frequency $\nuzb\approx 114$ MHz. 
For $B=5.3$ T, the trap-modified  cyclotron and anomaly frequencies are $\nucb \approx 149$ GHz and $\nuab \approx 173$ MHz \cite{Review}.  A circular magnetron motion at $\numb = 43$ kHz is cooled by axial sideband cooling  \cite{VanDyckMagnetronCoolingLimit,Review} and its effect is negligible during the measurement. 
Figure~\ref{fig:ApparatusAndLevels}c shows the lowest cyclotron and spin energy levels and the frequency spacings, including a relativistic mass shift, $\delta$, given by $\delta /\nu_c \equiv h\nu_c/(mc^2) \approx 10^{-9}$ \cite{Review,Gabrielse85e_RelativisticMassBistableHysteresis}. 

The lowest cyclotron states for each spin are effectively stable because the spin is so nearly uncoupled from its environment \cite{Review}.  Without a trap, the cyclotron state $n=1$ has a lifetime $\gamma_c^{-1}=0.1$ s. With a trap that is also a low-loss microwave cavity, this rate for the spontaneous emission of synchrotron radiation is inhibited by a factor of 50 to 70. Cyclotron excitations can then be detected before decay, when $B$ is chosen so $\nucb$ is far from resonance with cavity radiation modes \cite{InhibitionLetter}. The cyclotron damping contributes 0.03 Hz to the cyclotron and anomaly linewidths (to be discussed), a negligible 0.2 ppt and a very important 0.2 ppb, respectively. 
Blackbody photons that excite $n=0$ to $n=1$ are eliminated by cooling the trap cavity below 100 mK \cite{QuantumCyclotron}. 

The Brown-Gabrielse invariance theorem \cite{InvarianceTheorem},
\begin{equation}
\nu_c = \sqrt{\nucb^2 + \nuzb^2 + \numb^2}
\label{eq:InvarianceTheorem}
\end{equation}
provides the $\nu_c$ and $\nu_a = \nu_s-\nu_c$ needed in Eq.~\eqref{eq:FreeSpacegOver2} to determine $\mu/\mu_B$.
It is critical that Eq.~\eqref{eq:InvarianceTheorem} is invariant under unavoidable misalignments of {\bf B} and the axis of $V$, and under elliptic distortions of $V$. The hierarchy $\nucb \gg \nuzb \gg \numb \gg \delta$ allows an expansion of Eq.~\eqref{eq:InvarianceTheorem} that suffices for our precision to be inserted in Eq.~\eqref{eq:FreeSpacegOver2}, so
\begin{equation}
-\frac{\mu}{\mu_B} 
=\frac{g}{2} 
\simeq 1 + \frac{\nuab - \nuzb^2/(2 \fcb)}{\fcb + 3\delta/2 + \nuzb^2/(2 \fcb)} + \frac{\Delta g_{cav}}{2}, \label{eq:Experimentalg}
\end{equation} 
with $\nuab$ and $\fcb$  (defined in Fig.~\ref{fig:ApparatusAndLevels}c) to be deduced with $\nuzb$ from measured line shapes.
The added cavity-shift $\Delta g_{cav}/2$ arises because the cyclotron frequency couples to radiation modes of the trap cavity, shifting both $\nucb$ and $\nuab$ \cite{RenormalizedModesPRL,RenormalizedModesPRA}. This $g/2$ measurement correction and its uncertainty are not reduced or evaded by a $g-2$ measurement.  They must be determined and corrected at the full $10^{-13}$ precision of $\mu/\mu_B$.

\begin{figure}[htbp!]
\centering
\includegraphics[width=\the\columnwidth]{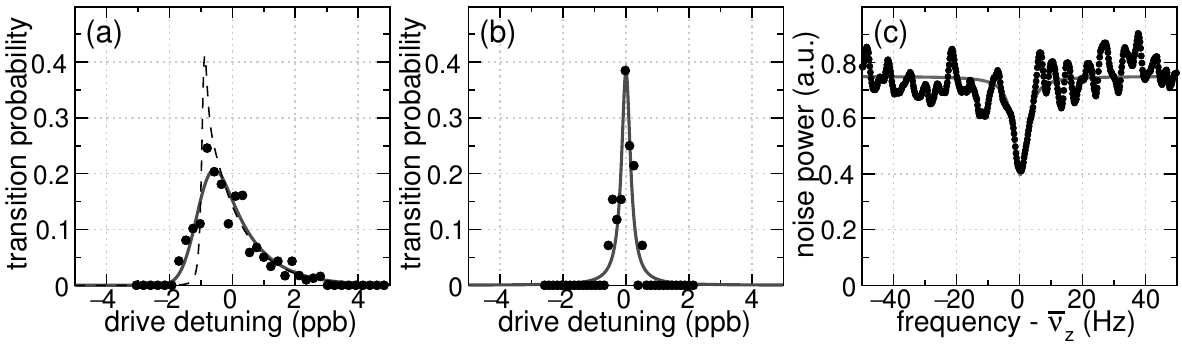}
\caption{Quantum jump cyclotron (a) and anomaly (b) line shapes that are measured (points), predicted (dashed) and fit (solid) vs fractional drive detunings from $\fcb (1+\epsilon)$ and $\nuab (1+\epsilon)$ (defined later in the text). (c) A dip in Johnson noise reveals $\nuzb$.  }
\label{fig:Lineshapes}
\end{figure}

To measure the $\nuzb$ needed in Eq.~\eqref{eq:Experimentalg}, the current induced in the electrodes by the axial oscillation is sent through a resonant circuit that is the input of a cryogenic HEMT amplifier. The 1-minute Fourier transform of the amplifier output in Fig.~\ref{fig:Lineshapes}c illustrates the Johnson noise and electron signal canceling to make a dip that reveals $\nuzb$  \cite{ElectronCalorimeter}.  
Energy loss in the circuit damps the axial motion with a time constant $\gamma_z^{-1}=32$ ms.  The amplifier heats the electron axial motion to $T_z=0.5$ K.    

Small shifts in $\nuzb$ provide quantum nondemolition detection (QND) of one-quantum spin and cyclotron jumps, without the detection changing the cyclotron or spin state.  Saturated nickel rings (Fig.~\ref{fig:ApparatusAndLevels}b) produce a  magnetic bottle gradient, $\Delta B=B_2\left[(z^2 - \rho^2/2)\hat{z}-z\rho\hat{\rho}\right]$ with $B_2 = 300~\rm{T}/\rm{m}^2$.  This couples spin and cyclotron energies to $\nuzb$, which then shifts by $\Delta \nuzb \approx 1.3~(n + m_s)$ Hz.  (The $B_2$ and $\Delta \nuzb$ are 5 and 3 times smaller than used previously  \cite{HarvardMagneticMoment2008}.) 
To rapidly detect jumps after the cyclotron and anomaly drives are turned off, the amplified signal is immediately fed back to the electron. This self-excited oscillator (SEO) \cite{SelfExcitedOscillator}   resonantly and rapidly drives itself to a large amplitude even if $\nuzb$ shifts with amplitude, whereupon the gain is adjusted to maintain the amplitude. A Fourier transform of the large signal reveals the small $\Delta \nuzb$ that signals cyclotron and spin jumps.

Quantum jump spectroscopy produces anomaly and cyclotron resonances (Fig.~\ref{fig:Lineshapes}a-b) from which to extract $\nuab$ and $\fcb$ to use in Eq.~\eqref{eq:Experimentalg}. Cyclotron and anomaly quantum jump trials are alternated.  The magnetic field drift of 0.2~ppb/hr in the new apparatus is slow enough to correct using a quadratic fit to the lowest cyclotron drive frequencies that produce excitations. Each cyclotron and anomaly quantum jump trial starts with resonant anomaly and cyclotron drives that prepare the electron in the spin-up ground state, $\left|n=0,m_s=1/2\right\rangle$, followed by 1~s of axial magnetron sideband cooling \cite{VanDyckMagnetronCoolingLimit,Review}. 

Cyclotron jumps to $n=1$ are driven by a 5 s microwave drive injected between trap electrodes (Fig.~\ref{fig:ApparatusAndLevels}b), with an off-resonance anomaly drive also applied.  Jumps occur in less than 20\% of the trials to avoid saturation effects.  Cavity-inhibited spontaneous emission \cite{InhibitionLetter} makes the excitation persist long enough so that  self-excitation feedback \cite{SelfExcitedOscillator} can be turned on in the next 1~s to detect the 1.3 Hz shift that signals a cyclotron quantum jump. 

Anomaly quantum jumps are driven by an oscillatory potential applied to trap electrodes for 30 s to drive an off-resonance axial oscillation of the electron through the radial magnetic gradient $B_2z \boldsymbol{\rho}$.  A cyclotron drive remains applied but is off resonance.
The electron sees the oscillating magnetic field perpendicular to $\hat{z}$ as needed to flip its spin, with a radial gradient that allows a simultaneous cyclotron transition \cite{Review}.
A spontaneous decay to the spin-down ground state, $\left|n=0,m_s=-1/2\right\rangle$, would be detected during the 60 s (more than 10 cyclotron decay times) after the drives are turned off.
A maximum jump rate of 40\% suggests a slight power broadening, but $\nuab$ is still determined far more precisely than $\fcb$.

Well-understood, asymmetric cyclotron and symmetric anomaly line shapes are predicted \cite{BrownLineshape} for thermal axial motion at temperature $T_z$ within a magnetic gradient $B_2 z^2$.  To this, the effect of cyclotron decay has been added \cite{HarvardMagneticMoment2011}.
The average oscillation amplitude squared  is $\overline{z^2} =  k_B T_z/(4 \pi^2\nuzb^2m)$, where $k_B$ is the Boltzmann constant. The average field for the electron is shifted by $\epsilon B = B_2 \overline{z^2}$ and broadened by the same amount.  The cyclotron bandwidth $\epsilon \nucb$ corresponds to a time $(\epsilon \nucb )^{-1} = 1.3$ ms needed to establish $\nucb$.  This is much faster than the $\gamma_z^{-1} = 32$ ms scale on which the axial amplitude fluctuates, so the predicted cyclotron line shape (dashed in Fig.~\ref{fig:Lineshapes}a)  approximates an exponential Boltzmann shape, centered at frequency $\nucb (1+\epsilon)$.  The anomaly transition time $(\epsilon \nuab)^{-1} = 1.1$ s is much slower than the axial amplitude fluctuations, whereupon the predicted thermal anomaly line is essentially symmetric about $\nuab (1+\epsilon)$ and is negligibly narrow. 
The observed anomaly linewidth of 0.06 Hz (0.35 ppb) in Fig.~\ref{fig:Lineshapes}b is from other sources. Half is from the cyclotron decay lifetime and half is from applying the anomaly drive for only 30 s.

The anomaly line shape is consistent with what is predicted but the cyclotron line shape is not.  Presumably this is due to unwanted magnetic field fluctuations that are averaged differently in the anomaly and cyclotron line shapes.  Such fluctuations, with a 200 Hz bandwidth, were observed with a superconducting solenoid being jostled by its environment \cite{BollingerFieldFluctuations}. The anomaly line shape would average away such fluctuations to yield the narrow line observed (e.g.\ Fig.~\ref{fig:Lineshapes}b). The cyclotron line shape would not, giving a possible explanation for the observed  0.5--0.8 ppb broadening (e.g.\ Fig.~\ref{fig:Lineshapes}a).

Both $\nuab$ and $\fcb$ are extracted from such line shapes.  Cyclotron line shapes are fit to the predicted line shape (dashed in Fig.~\ref{fig:Lineshapes}a),  convoluted with a Gaussian function to accommodate the broadening. Such a fit, illustrated by the solid curve in Fig.~\ref{fig:Lineshapes}a, typically gives a 2 ppb cyclotron linewidth, a Gaussian broadening width of about 0.5 ppb, $T_z = 0.55 \pm 0.11$ K, and $\fcb$ with an uncertainty of about $0.08$ ppb.   
For anomaly line shapes (e.g.\  Fig.~\ref{fig:Lineshapes}b), nearly symmetric and fractionally narrower by about a factor of 4, the uncertainty in $\nuab$ is thus not very significant for the final uncertainty.  Fitting with or without Gaussian broadening makes little difference
(e.g.\ solid curve in Fig.~\ref{fig:Lineshapes}b).

\begin{figure}[htbp!]
\centering
    \includegraphics[width=\the\columnwidth]{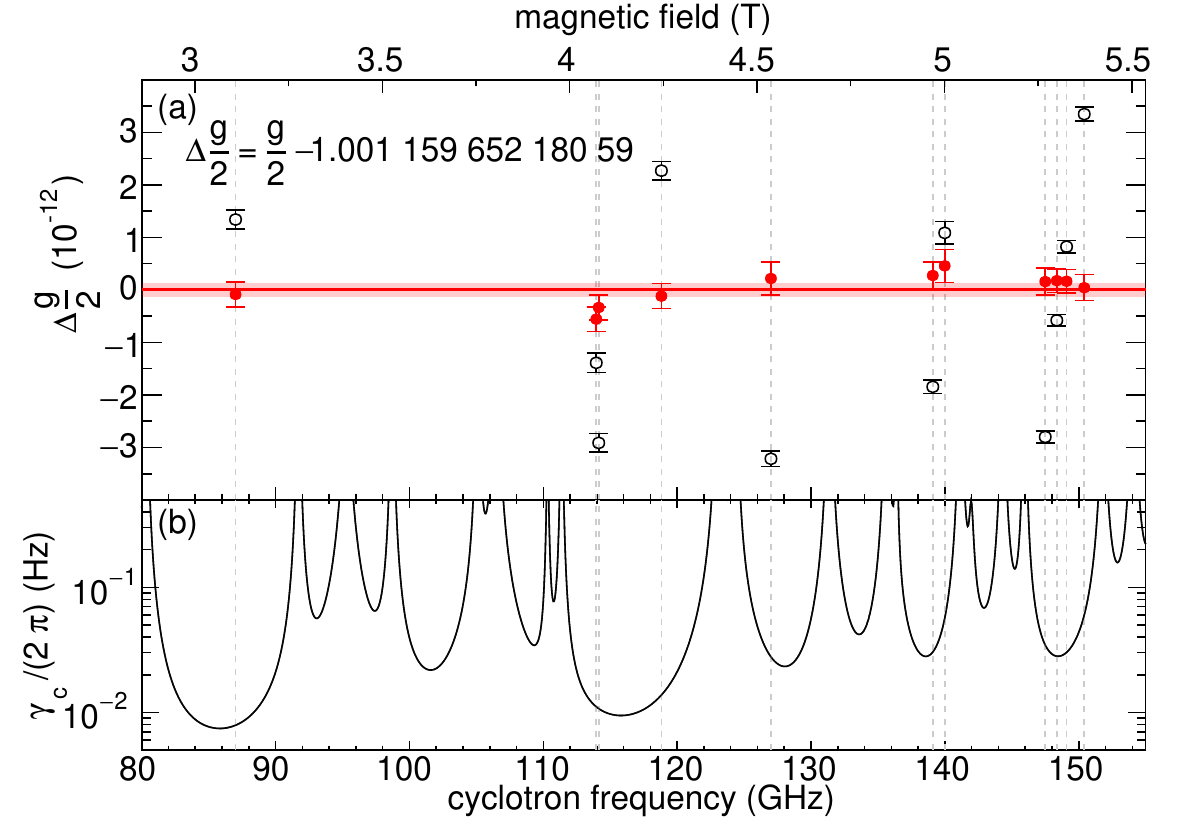}
 \caption{(a) Measured $g/2-1.001\,159\,652\,180\,59$ before (white) and after (red) cavity-shift correction. (b) Measurements take place in valleys of the  cyclotron damping rate where spontaneous emission is inhibited.}
   \label{fig:ManyBFields}
\end{figure}  

The cavity-shift $\Delta g_{cav}/2$ in Eq.~\eqref{eq:Experimentalg}, the only correction to what is directly measured, arises because the cyclotron oscillator couples to radiation modes of the trap cavity  and shifts $\nucb$ \cite{RenormalizedModesPRA,RenormalizedModesPRL}.  It is the downside of the cavity-inhibited spontaneous emission that desirably narrows resonance lines, and makes it possible to observe a cyclotron excitation before it decays. The cylindrical trap was invented \cite{CylindricalPenningTrap} to allow cavity modes and shifts to be understood and calculated. Nonetheless, the  mode frequencies and $Q$ values must still be measured because of energy losses in induced surface currents,  imperfect cavity machining, slits that make cavity sections into separately-biased trap electrodes, and  dimension changes as the cavity cools below 100 mK from 300 K.  Three consistent methods are used: (1) parametrically-pumped electrons \cite{SynchronizedElectronsPRA,SynchronizedElectronsPRL,HarvardMagneticMoment2011}, (2) measuring how long one electron stays in its first excited cyclotron state \cite{HarvardMagneticMoment2008,HarvardMagneticMoment2011}, and (3) a new method of observing the decay time of an electron exited to $n_c \approx 10$.

A renormalized calculation \cite{RenormalizedModesPRA,RenormalizedModesPRL} with added cyclotron damping \cite{HarvardMagneticMoment2008,HarvardMagneticMoment2011} avoids the infinite cavity shifts that result from summing all mode contributions. This calculation assumes the mode frequencies of a perfect cylinder, one $Q$ for TE modes, and another for TM modes.  We calculate with  dimensions chosen to best match observed frequencies and  a single $Q$ value for all modes.  After shifts from the 72 observed modes using the ideal frequencies and the one $Q$ value are subtracted out, contributions for these modes using measured frequencies and $Q$ values are added back in. The leading contribution to cavity shift uncertainties comes from modifications of the field that an electron sees from imperfections and misalignments of the trap cavity. Figure~\ref{fig:ManyBFields}a shows the consistency of $\mu/\mu_B$ determinations at 11 different magnetic fields, after each receives a different cavity shift.

A weighted average of the 11 determinations gives        
\begin{equation}
-\frac{\mu}{\mu_B} = \frac{g}{2} = 1.001 \, 159 \, 652 \, 180\, 59\,(13)~~[0.13~\rm{ppt}], 
\end{equation}  
with 1$\sigma$ uncertainty in the last two digits in parentheses.  
Figure~\ref{fig:Result} shows the good agreement of this 2022 determination at Northwestern with our 2008 determination at Harvard \cite{HarvardMagneticMoment2008} and an uncertainty that is improved by a factor of 2.2. 
Because uncertainty correlations from similar measurement methods are difficult to determine, we do not recommend averaging our two determinations. 
 Table~\ref{table:Uncertainties} lists uncertainty contributions to the final result. The statistical uncertainty is from the fits that extract $\fcb$ and $\nuab$.  The two dominant uncertainties have been discussed -- cyclotron broadening and cavity shifts (treated as correlated for nearby fields).  The nuclear paramagnetism uncertainty is based upon the measured  temperature fluctuations of the silver trap electrodes.  The anomaly power shift uncertainty comes from the measured frequency dependence on drive strength.

\begin{table}[htbp!]
\centering
\caption{Largest uncertainties for $g/2$.}
\begin{tabular}{c c} 
 \hline
 {\bf Source} & {\bf Uncertainty}$\times 10^{13}$\\ 
 \hline\hline
 statistical & 0.29\\
\hline
 cyclotron broadening& 0.94\\
cavity correction& 0.90\\
 nuclear paramagnetism & 0.12\\ 
 anomaly power shift  & 0.10\\
   magnetic field drift  & 0.09\\
\hline
total  & 1.3\\
\hline
\end{tabular}
   \label{table:Uncertainties}
\end{table}

\begin{figure}[htbp!]
\centering
 \includegraphics[width=\the\columnwidth]{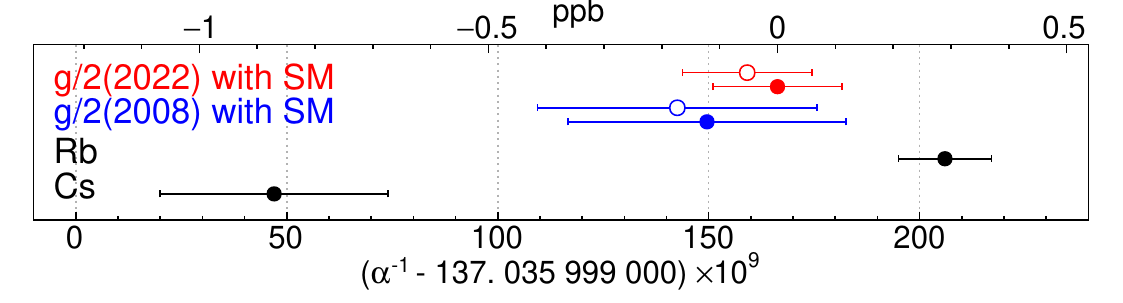}
   \caption{SM prediction of $\alpha$ using $\mu/\mu_B$ from this Northwestern measurement (red), and from our 2008 Harvard measurement (blue), with solid and open points for slightly differing  $C_{10}$ \cite{atomsTheoryReview2019,Volkov2019}. The    $\alpha$ measurements (black) were made with Cs at Berkeley \cite{MullerAlpha2018} and Rb in Paris \cite{RbAlpha2020}. A ppb is $10^{-9}$.
   }  
   \label{fig:ResultAlpha}
\end{figure}

Several SM sectors together predict
\begin{eqnarray}
\frac{g}{2}= &1& + \, C_2\left(\frac{\alpha}{\pi}\right) +C_4\left(\frac{\alpha}{\pi}\right)^2
+C_6\left(\frac{\alpha}{\pi}\right)^3
+C_8\left(\frac{\alpha}{\pi}\right)^4 \nonumber\\
&+&C_{10}\left(\frac{\alpha}{\pi}\right)^5 + ... + a_{\mu\tau} + a_{\rm{hadronic}} + a_{\rm{weak}}.
\label{eq:SMTheory}
\end{eqnarray}
The Dirac prediction \cite{DiracTheoryOriginal} is first on the right. 
QED provides the asymptotic series in powers of $\alpha$, along with the muon and tauon contributions $a_{\mu\tau}$  \cite{atomsTheoryReview2019}. The constants $C_2$ \cite{Schwinger}, $C_4$ \cite{Petermann,Sommerfield:1958}, $C_6$ \cite{Kinoshita:1995,LaportaRemiddi:1996} and $C_8$ \cite{QED_C8_Lapo} are calculated exactly, but require measured lepton mass ratios as input \cite{KURZ2014_HeavyLeptons}. 
The measurements are so precise that a numerically calculated tenth order $C_{10}$ \cite{TenthOrderQED2012,QED_C10_nio} is required and tested.  A second evaluation of $C_{10}$ \cite{Volkov2019} differs slightly for reasons not yet understood and the open points in Figs.~\ref{fig:Result} and \ref{fig:ResultAlpha} use this alternative.   
Hadronic and weak interaction contributions are $a_{\rm{hadronic}}$ \cite{Nomura2013,HadronicContribution2014,WeakHadronic} and $a_{\rm{weak}}$ \cite{WeakCondtribution1,WeakCondtribution2,WeakCondtribution3,WeakCondtribution4}.
The exact $C_8$ and the numerical $C_{10}$ are remarkable advances  that reduce the calculation uncertainty well below the uncertainties reported for the measured $\mu/\mu_B$ and $\alpha$.

The most precise $\alpha$ measurements  \cite{MullerAlpha2018,RbAlpha2020}, needed for the SM prediction of $g/2$ in  Eq.~\eqref{eq:SMTheory}, disagree by 5.5 $\sigma$, about ten times our measurement uncertainty (Fig.~\ref{fig:Result}).
Until the discrepancy is resolved, the best that can be said is that the predicted and measured $\mu/\mu_B$ agree to about $\delta (g/2) = 0.7\times10^{-12}$, half of the  $\alpha$ discrepancy.  A generic chiral symmetry model \cite{BrodskyDrellElectronSubstructure} then suggests that the electron radius is less than $R_e = \sqrt{|\delta (g/2)|}\hbar/(mc) = 3.2 \times 10^{-19}$~m, and that the mass of possible electron constituents must exceed $m^* = m/\sqrt{|\delta (g/2)|} = 620$~GeV/$c^2$.
If $\delta (g/2)$ would equal our $\mu/\mu_B$ determination uncertainty, then $R_e = 1.4 \times 10^{-19}$~m and $m^*=1.4$~TeV/$c^2$.

A 2.2 times reduced $\delta (g/2)$ would bring us to the level of the intriguing $4.2$ standard deviation discrepancy between the measured and predicted muon magnetic moment \cite{Muong-2PRL_2021,MuonAnomalyCalculation2020}. The muon's BSM sensitivity, expected to be 40,000 times higher (the ratio of muon and electron masses), is largely offset by our 3150 times smaller uncertainty.

The fine structure constant $\alpha$ is the fundamental measure of the strength of the electromagnetic interaction in the low energy limit.  For the SI system of units, $\alpha = e^2/(4\pi\epsilon_0 \hbar c)$ is a measure of the vacuum permittivity $\epsilon_0$, given that 
and $e$, $\hbar$ and the speed of light $c$ are now defined \cite{RedefineSI_2019}.
Our $\mu/\mu_B$ and the SM give  
\begin{eqnarray}
\alpha^{-1} &=& 137.035 \, 999 \, 166 \, (02)\,(15)~~[0.014~\rm{ppb}]\,[0.11~\rm{ppb}],\nonumber \\
&=& 137.035 \, 999 \, 166 \, (15)~~~~~~~~[0.11~\rm{ppb}], \label{eq:AlphaValue}
\end{eqnarray}
with  theoretical and experimental uncertainties in the first and second brackets. 
Figure~\ref{fig:ResultAlpha} compares to the $\alpha$ measurements (black) that disagree with each other by 5.5 $\sigma$. Our value differs by 2.1 standard deviations from the Paris Rb determination of $\alpha$ \cite{RbAlpha2020} and by 3.9 standard deviations from the Berkeley Cs determination \cite{MullerAlpha2018}.
The $C_{10}$ in \cite{Volkov2019} would change only ``66'' to ``59'' in Eq.~\eqref{eq:AlphaValue}.

For the future, a measurement is underway to realize the new precision with a positron, to improve the test of the fundamental $CPT$ symmetry invariance of the SM by a factor of 40 \cite{EfficientPositronAccumulation}.
Much larger improvements in the precision of $\mu/\mu_B$ now seem feasible given the  demonstration of more stable apparatus, improved statistics, and better understood uncertainties.
Detectors being tested, more harmonic and lower loss trap cavities, and detector backaction circumvention methods \cite{Fan2020BackActionPRL,Fan2020BackActionPRA} should enable  much more precise  measurements to come.   

In conclusion, an electron magnetic moment measurement is carried out blind to previous measurements and predictions. A PhD thesis \cite{ThesisFan} and a longer publication in preparation give fuller accounts.  In new apparatus at a different university, the measured $\mu/\mu_B$ is consistent with our 2008 measurement, with a factor of 2.2 improved precision.  The most precise prediction of the SM agrees with the most precise determination of a property of an elementary particle to about $1$ part in $10^{12}$. When  discrepant $\alpha$ measurements are resolved, the new measurement uncertainty of $1.3$ parts in $10^{13}$ is available for a more precise test for BSM physics 

\begin{acknowledgments}
Early contributions were made by S.\ E.\ Fayer. 
NSF 2110565 provided the support, with X.\ Fan supported by the Masason Foundation.  Detector development is supported by the Templeton Foundation, and low-loss trap cavity development is supported by the DOE SQMS Center.   
\end{acknowledgments}
 
\bibliographystyle{prsty_gg}
\bibliography{ggrefs2018,NewRefs}

\begin{thebibliography}{10}

\bibitem{Sakharov}
A.~D. {Sakharov}, Soviet Journal of Experimental and Theoretical Physics
  Letters {\bf 5},  24  (1967).

\bibitem{ExpandingUniverse1946}
G. Gamow, Phys. Rev. {\bf 70},  572  (1946).

\bibitem{DMDiscovery1933}
F. {Zwicky}, Helvetica Physica Acta {\bf 6},  110  (1933).

\bibitem{DMRotationCurve1980}
V.~C. {Rubin}, J. {Ford}, W.~K., and N. {Thonnard}, \apj {\bf 238},  471
  (1980).

\bibitem{Perlmutter_1999_AcceleratingUniverse}
S. Perlmutter {\it et~al.}, The Astrophysical Journal {\bf 517},  565  (1999).

\bibitem{Riess_1998_AcceleratingUniverse}
A.~G. Riess {\it et~al.}, The Astronomical Journal {\bf 116},  1009  (1998).

\bibitem{Linde1982Inflation}
A. Linde, Physics Letters B {\bf 108},  389  (1982).

\bibitem{STAROBINSKY1980Inflation}
A. Starobinsky, Physics Letters B {\bf 91},  99  (1980).

\bibitem{STAROBINSKY1982Inflation}
A. Starobinsky, Physics Letters B {\bf 117},  175  (1982).

\bibitem{Guth1981Inflation}
A.~H. Guth, Phys. Rev. D {\bf 23},  347  (1981).

\bibitem{ScreenedFifthForceLimitFromGFactor2018}
P. Brax, A.-C. Davis, B. Elder, and L.~K. Wong, Phys. Rev. D {\bf 97},  084050
  (2018).

\bibitem{HiggsDoubletCPViolation2019}
E.~J. Chun, J. Kim, and T. Mondal, Journal of High Energy Physics {\bf 2019},
  68  (2019).

\bibitem{Giudice2012}
G.~F. Giudice, P. Paradisi, and M. Passera, Journal of High Energy Physics {\bf
  2012},  113  (2012).

\bibitem{ProbeNewPhysicsWithElectronG2014}
A. Aboubrahim, T. Ibrahim, and P. Nath, Phys. Rev. D {\bf 89},  093016  (2014).

\bibitem{HiddenPhotonConstraintsHamaguchi}
M. Endo, K. Hamaguchi, and G. Mishima, Phys. Rev. D {\bf 86},  095029  (2012).

\bibitem{PhysRevD.98.075011}
H. Davoudiasl and W.~J. Marciano, Phys. Rev. D {\bf 98},  075011  (2018).

\bibitem{ALPExplanationOfAnomaly2020}
M. Bauer, M. Neubert, S. Renner, M. Schnubel, and A. Thamm, Phys. Rev. Lett.
  {\bf 124},  211803  (2020).

\bibitem{ALightComplexScalarForTheElectronAndMuonAnomalousMagneticMoments}
J. Liu, C.~E.~M. Wagner, and X.-P. Wang, Journal of High Energy Physics {\bf
  2019},  8  (2019).

\bibitem{SUSYSenarioForElectronAndMuong_2019}
M. Endo and W. Yin, Journal of High Energy Physics {\bf 2019},  122  (2019).

\bibitem{DiracTheoryOriginal}
P.~A.~M. {Dirac}, Proceedings of the Royal Society of London Series A {\bf
  118},  351  (1928).

\bibitem{Schwinger}
J. Schwinger, Phys. Rev. {\bf 73},  416  (1948).

\bibitem{Petermann}
A. Petermann, Helv. Phys. Acta {\bf 30},  407  (1957).

\bibitem{Sommerfield:1958}
C.~M. Sommerfield, {Ann.\ Phys.\ (N.Y.)} {\bf 5},  26  (1958).

\bibitem{Kinoshita:1995}
T. Kinoshita, Phys. Rev. Lett. {\bf 75},  4728  (1995).

\bibitem{LaportaRemiddi:1996}
S. Laporta and E. Remiddi, Phys. Lett. B {\bf 379},  283  (1996).

\bibitem{QED_C8_Lapo}
S. Laporta, Physics Letters B {\bf 772},  232   (2017).

\bibitem{QED_C10_nio}
T. Aoyama, T. Kinoshita, and M. Nio, Phys. Rev. D {\bf 97},  036001  (2018).

\bibitem{TenthOrderQED2012}
T. Aoyama, M. Hayakawa, T. Kinoshita, and M. Nio, Phys. Rev. Lett. {\bf 109},
  111807  (2012).

\bibitem{KURZ2014_HeavyLeptons}
A. Kurz, T. Liu, P. Marquard, and M. Steinhauser, Nuclear Physics B {\bf 879},
  1  (2014).

\bibitem{Nomura2013}
D. Nomura and T. Teubner, Nucl. Phys. B {\bf 867},  236  (2013).

\bibitem{HadronicContribution2014}
A. Kurz, T. Liu, P. Marquard, and M. Steinhauser, Physics Letters B {\bf 734},
  144   (2014).

\bibitem{WeakHadronic}
{Jegerlehner, Fred}, EPJ Web Conf. {\bf 218},  01003  (2019).

\bibitem{WeakCondtribution1}
K. Fujikawa, B.~W. Lee, and A.~I. Sanda, Phys. Rev. D {\bf 6},  2923  (1972).

\bibitem{WeakCondtribution2}
A. Czarnecki, B. Krause, and W.~J. Marciano, Phys. Rev. Lett. {\bf 76},  3267
  (1996).

\bibitem{WeakCondtribution3}
M. Knecht, M. Perrottet, E. de~Rafael, and S. Peris, Journal of High Energy
  Physics {\bf 11},  003  (2002).

\bibitem{WeakCondtribution4}
A. Czarnecki, W.~J. Marciano, and A. Vainshtein, Phys. Rev. D {\bf 67},  073006
   (2003).

\bibitem{HarvardMagneticMoment2008}
D. Hanneke, S. Fogwell, and G. Gabrielse, Phys. Rev. Lett. {\bf 100},  120801
  (2008).

\bibitem{atomsTheoryReview2019}
T. {Aoyama}, T. {Kinoshita}, and M. {Nio}, Atoms {\bf 7},  28  (2019).

\bibitem{Volkov2019}
S. Volkov, Phys. Rev. D {\bf 100},  096004  (2019).

\bibitem{MullerAlpha2018}
R.~H. Parker, C. Yu, W. Zhong, B. Estey, and H. M{\"u}ller, Science {\bf 360},
  191  (2018).

\bibitem{RbAlpha2020}
L. Morel, Z. Yao, P. Clad{\'e}, and S. Guellati-Kh{\'e}lifa, Nature {\bf 588},
  61  (2020).

\bibitem{QuantumCyclotron}
S. Peil and G. Gabrielse, Phys. Rev. Lett. {\bf 83},  1287  (1999).

\bibitem{Muong-2PRL_2021}
B. Abi {\it et~al.}, Phys. Rev. Lett. {\bf 126},  141801  (2021).

\bibitem{Helium3NMR2019}
X. Fan, S.~E. Fayer, and G. Gabrielse, Review of Scientific Instruments {\bf
  90},  083107  (2019).

\bibitem{SelfShieldingSolenoid}
G. Gabrielse and J. Tan, J. Appl. Phys. {\bf 63},  5143  (1988).

\bibitem{Atoms2019TowardImprovedMeasurement}
G. Gabrielse, S. Fayer, T. Myers, and X. Fan, Atoms {\bf 7},  45  (2019).

\bibitem{Review}
L.~S. Brown and G. Gabrielse, Rev. Mod. Phys. {\bf 58},  233  (1986).

\bibitem{CylindricalPenningTrap}
G. Gabrielse and F.~C. MacKintosh, Intl. J. of Mass Spec. and Ion Proc. {\bf
  57},  1  (1984).

\bibitem{CylindricalPenningTrapDemonstrated}
J. Tan and G. Gabrielse, Appl. Phys. Lett. {\bf 55},  2144  (1989).

\bibitem{VanDyckMagnetronCoolingLimit}
R.~S. {Van Dyck, Jr.}, P.~B. Schwinberg, and H.~G. Dehmelt,  in {\em New
  Frontiers in High Energy Physics}, edited by B. Kursunoglu, A. Perlmutter,
  and L.Scott ({Plenum}, New York, 1978), p.\ 159.

\bibitem{Gabrielse85e_RelativisticMassBistableHysteresis}
G. Gabrielse, H. Dehmelt, and W. Kells, Phys. Rev. Lett. {\bf 54},  537
  (1985).

\bibitem{InhibitionLetter}
G. Gabrielse and H. Dehmelt, Phys. Rev. Lett. {\bf 55},  67  (1985).

\bibitem{InvarianceTheorem}
L.~S. Brown and G. Gabrielse, Phys. Rev. A {\bf 25},  2423  (1982).

\bibitem{RenormalizedModesPRL}
L.~S. Brown, G. Gabrielse, K. Helmerson, and J. Tan, Phys. Rev. Lett. {\bf 55},
   44  (1985).

\bibitem{RenormalizedModesPRA}
L.~S. Brown, G. Gabrielse, K. Helmerson, and J. Tan, Phys. Rev. A {\bf 32},
  3204  (1985).

\bibitem{ElectronCalorimeter}
D.~J. Wineland and H.~G. Dehmelt, J. Appl. Phys. {\bf 46},  919  (1975).

\bibitem{SelfExcitedOscillator}
B. D'Urso, R. {Van Handel}, B. Odom, D. Hanneke, and G. Gabrielse, Phys. Rev.
  Lett. {\bf 94},  113002  (2005).

\bibitem{BrownLineshape}
L.~S. Brown, {Ann.\ Phys.\ (N.Y.)} {\bf 159},  62  (1985).

\bibitem{HarvardMagneticMoment2011}
D. Hanneke, S. Fogwell~Hoogerheide, and G. Gabrielse, Phys. Rev. A {\bf 83},
  052122  (2011).

\bibitem{BollingerFieldFluctuations}
J.~W. Britton, J.~G. Bohnet, B.~C. Sawyer, H. Uys, M.~J. Biercuk, and J.~J.
  Bollinger, Phys. Rev. A {\bf 93},  062511  (2016).

\bibitem{SynchronizedElectronsPRA}
J. Tan and G. Gabrielse, Phys. Rev. A {\bf 48},  3105  (1993).

\bibitem{SynchronizedElectronsPRL}
J. Tan and G. Gabrielse, Phys. Rev. Lett. {\bf 67},  3090  (1991).

\bibitem{BrodskyDrellElectronSubstructure}
S.~J. Brodsky and S.~D. Drell, Phys. Rev. D {\bf 22},  2236   (1980).

\bibitem{MuonAnomalyCalculation2020}
T. Aoyama {\it et~al.}, Physics Reports {\bf 887},  1  (2020).

\bibitem{RedefineSI_2019}
M. Stock, R. Davis, E. de~Mirand{\'{e}}s, and M.~J.~T. Milton, Metrologia {\bf
  56},  022001  (2019).

\bibitem{EfficientPositronAccumulation}
S. Fogwell~Hoogerheide, J.~C. Dorr, E. Novitski, and G. Gabrielse, Review of
  Scientific Instruments {\bf 86},  053301  (2015).

\bibitem{Fan2020BackActionPRL}
X. Fan and G. Gabrielse, Phys. Rev. Lett. {\bf 126},  070402  (2021).

\bibitem{Fan2020BackActionPRA}
X. Fan and G. Gabrielse, Phys. Rev. A {\bf 103},  022824  (2021).

\bibitem{ThesisFan}
X. Fan, Ph.D. thesis, Harvard University, 2022, (thesis advisor: G. Gabrielse).

\end{thebibliography}

\end{document}